\documentclass[aps,prb,twocolumn,showpacs]{revtex4}

\usepackage{graphicx}
\usepackage{amsmath}
\usepackage{amssymb}

\begin{document}

\title{Peak-structure in self-energy of cuprate superconductors}

\author{Yiqun Liu$^{1}$\footnotemark[1], Yu Lan$^{2}$\footnotemark[1]\footnotetext[1]{These authors
contributed equally to this work}, and Shiping Feng$^{1}$\footnote[2]{E-mail address: spfeng@bnu.edu.cn}}
%\email{spfeng@bnu.edu.cn}

\affiliation{$^{1}$Department of Physics, Beijing Normal University, Beijing 100875, China}

\affiliation{$^{2}$College of Physics and Electronic Engineering, Hengyang Normal University, Hengyang
421002, China}

\begin{abstract}
The recently deduced normal and anomalous self-energies from photoemission spectra of cuprate
superconductors via the machine learning technique are calling for an explanation. Here the normal and
anomalous self-energies in cuprate superconductors are analyzed within the framework of the
kinetic-energy-driven superconductivity. It is shown that the exchanged spin excitations give rise to
the well-pronounced low-energy peak-structures in both the normal and anomalous self-energies, however,
they do not cancel in the total self-energy. In particular, the peak-structure in the normal self-energy
is mainly responsible for the peak-dip-hump structure in the single-particle excitation spectrum, and
can persist into the normal-state, while the sharp peak in the anomalous self-energy gives rise to a
crucial contribution to the superconducting gap, and vanishes in the normal-state. Moreover, the
evolution of the peak-structure with doping and momentum are also analyzed.
\end{abstract}


\maketitle

\section{Introduction}\label{Introduction}

The strong electron correlation is foundational to the emergence of superconductivity in cuprate
superconductors \cite{Bednorz86,Anderson87}, where the strong interaction of the electrons with
collective bosonic excitations of different origins results in (i) the energy and lifetime
renormalization of the electrons in the particle-hole channel to form the quasiparticles responsible
for the anomalous properties, and (ii) the formation of the electron pairs in the particle-particle
channel responsible for superconductivity below the superconducting (SC) transition temperature
$T_{\rm c}$. This is why the quasiparticles in the SC-state determined by the electronic structure is
intimately related to the pairing glue forming electron pairs
\cite{Damascelli03,Campuzano04,Fink07,Carbotte11,Bok16}. In conventional superconductors, the
renormalization of the electrons in the particle-hole channel and the formation of the electron pairs
in the particle-particle channel are caused by the interaction between the electrons by the exchange
of phonons \cite{Bardeen57,Eliashberg60,Scalapino66}. Although a significant effort has been made for
the past three decades, what type of the collective bosonic excitation that can mediate electron
pairing in cuprate superconductors in analogy to the phonon-mediate pairing mechanism in conventional
superconductors is still debated \cite{Anderson87,Damascelli03,Campuzano04,Fink07,Carbotte11,Bok16}.

Angle-resolved photoemission spectroscopy (ARPES) is a direct tool to probe the energy and momentum of
quasiparticles simultaneously \cite{Damascelli03,Campuzano04,Fink07}, where a quasiparticle with a
long lifetime is observed as a sharp peak in intensity, and a quasiparticle with a short lifetime is
observed as a broad peak. However, the energy and lifetime of the quasiparticle in the SC-state are
directly described by the real and imaginary parts of the total self-energy
\cite{Damascelli03,Campuzano04,Fink07,Carbotte11,Bok16} ${\rm Re}\Sigma_{\rm tot}({\bf k},\omega)$
and ${\rm Im}\Sigma_{\rm tot}({\bf k},\omega)$, respectively. This total self-energy
$\Sigma_{\rm tot}({\bf k},\omega)$ is a specific combination of the normal self-energy
$\Sigma_{\rm ph}({\bf k},\omega)$ in the particle-hole channel and the anomalous self-energy
$\Sigma_{\rm pp}({\bf k},\omega)$ in the particle-particle channel
\cite{Eliashberg60,Scalapino66,Gorkov58,Nambu60}. In other words, only the total self-energy can be
extracted directly from ARPES experiments, and the only ingredient that needs to extract the total
self-energy is the quasiparticle spectral density observed by ARPES experiments
\cite{Damascelli03,Campuzano04,Fink07}. However, for our exploration of the bosonic mode coupling that
is how electron self-energy effects appeared in our theoretical analysis, it is crucial to extract the
normal and the anomalous self-energies separately \cite{Eliashberg60,Scalapino66,Gorkov58,Nambu60}.
This follows a basic fact that although both the normal and anomalous self-energies are generated by
the same electron interaction mediated by collective bosonic excitations, they describe theoretically
different parts of the interaction effects. The normal self-energy $\Sigma_{\rm ph}({\bf k},\omega)$
describes the single-particle coherence, and therefore competes with superconductivity. Moreover, it
gives rise to a main contribution to the energy and lifetime renormalization of the electrons, and then
all the anomalous properties of cuprate superconductors arise from this renormalization of the electrons
\cite{Timusk99,Hufner08,Comin16,Vishik18}. On the other hand, the SC-state is characterized by the
anomalous self-energy $\Sigma_{\rm pp}({\bf k},\omega)$, which is identified as the energy and momentum
dependent SC gap in the single-particle excitation spectrum, and therefore is corresponding to the
energy for breaking an electron pair \cite{Eliashberg60,Scalapino66,Gorkov58,Nambu60}. In this case, if
both the normal and anomalous self-energies are deduced from the experimental data, it can be used to
examine a microscopic SC theory and understand the details of the SC-state.

Although both the normal and anomalous self-energies can not be measured directly from ARPES experiments,
the Boltzmann-machine learning technique \cite{Ackley85}  has been applied recently to deduce both the
normal and anomalous self-energies from the experimental data of the ARPES spectra observed in
Bi$_{2}$Sr$_{2}$CaCu$_{2}$O$_{8+\delta}$ at the optimum doping and Bi$_{2}$Sr$_{2}$CuO$_{6+\delta}$ in
the underdoped regime \cite{Yamaji19}, and the deduced results show clearly that both the normal and
anomalous self-energies exhibit the notable low-energy peak-structures, however, these low-energy
peak-structures do not appear in the total self-energy. In particular, the peak in the anomalous
self-energy makes a dominant contribution to the SC gap, and therefore provide a decisive testimony for
the origin of superconductivity \cite{Yamaji19}. These normal and anomalous self-energies of cuprate
superconductors revealed by the machine learning approach therefore are calling for a systematic analysis.
Quite recently, these deduced normal and anomalous self-energies in Ref. \onlinecite{Yamaji19} have been
analyzed within an effective fermion-boson theory \cite{Chubukov20}, and the result indicates that the
pairing electrons is mediated by a soft, near-critical bosonic mode. In particular, this analysis also
shows that if the sharp low-energy peaks in both the normal and anomalous self-energies survive down to
the lowest temperatures, their presence alone imposes the strong restrictions on the energy dependence
of a soft pairing boson \cite{Chubukov20}. This conclusion is also similar to that obtained in terms of
the machine learning approach \cite{Yamaji19}. However, the full understanding of these low-energy
peak-structures in both normal and anomalous self-energies is still open for further analyses. In this
paper, we make a comparison of the deduced normal and anomalous self-energies in
Ref. \onlinecite{Yamaji19} with those obtained based on the kinetic-energy-driven SC mechanism
\cite{Feng0306,Feng12,Feng15,Feng15a}, and then show explicitly that the interaction between electrons
by the exchange of a strongly dispersive spin excitation generates the sharp low-energy peak-structures
in both the normal and anomalous self-energies at around the antinodal region, which are in qualitative
agreement with the corresponding results in both the normal and anomalous self-energies deduced via the
machine learning technique \cite{Yamaji19}. However, these prominent low-energy peak-structures in both
the normal and anomalous self-energies do not cancel in the total self-energy. Although the absence of
this cancellation in the total self-energy is inconsistent with the corresponding result in the total
self-energy deduced from the machine learning method \cite{Yamaji19}, it is well consistent with the
corresponding experimental result in the total self-energy observed on cuprate superconductors
\cite{DMou17}. Moreover, we show clearly that the sharp low-energy peak-structure in the normal
self-energy is mainly responsible for the famous peak-dip-hump (PDH) structure in the single-particle
excitation spectrum \cite{Dessau91,Norman97,Campuzano99,Wei08,DMou17}, and can persist into the
normal-state, while the sharp low-energy peak in the anomalous self-energy gives rise to a crucial
contribution to the SC gap, and vanish in the normal-state.

In the remainder of this paper, the general formalism of the single-particle diagonal and off-diagonal
propagators (then the normal and anomalous self-energies) obtained within the framework of the
kinetic-energy-driven superconductivity is introduced briefly in Sec. \ref{Formalism}, while the
quantitative characteristics of the normal and anomalous self-energies and the related exotic features
of the electron quasiparticle excitations are presented in Sec. \ref{Quantitative-characteristics},
where we show that the sharp low-energy peak-structures in both the normal and anomalous self-energies
are doping dependent. In particular, in the underdoped regime, the position of the peak in the anomalous
self-energy at around the antinode moves further away from the Fermi energy with the increase of doping,
while the peak in the normal self-energy at around the $[\pi,0]$ point of the Brillouin zone (BZ) moves
towards to the Fermi energy. Furthermore, the sharp low-energy peak-structures also have a striking
momentum dependence, with the position of the peak in the normal self-energy that shifts towards to the
Fermi energy when one moves the momentum from the antinode to the node. Finally, we give a summary and
discussions in Sec. \ref{Conclusion}.

\section{Theoretical Framework}\label{Formalism}

In cuprate superconductors, the single common feature in the layered crystal structure is the presence
of the two-dimensional CuO$_{2}$ planes \cite{Bednorz86}, and then it seems evident that the
unconventional behaviors in cuprate superconductors are dominated by the strongly correlated motion of
the electrons in these CuO$_{2}$ planes. In this case, it has been suggested that the essential physics
of the doped CuO$_{2}$ plane can be properly accounted by the $t$-$J$ model on a square lattice
\cite{Anderson87},
\begin{eqnarray}\label{tjham}
H&=&-t\sum_{\langle ll'\rangle_{\rm NN}\sigma}C^{\dagger}_{l\sigma}C_{l'\sigma}
+t'\sum_{\langle ll'\rangle_{\rm NNN}\sigma}C^{\dagger}_{l\sigma}C_{l'\sigma} \nonumber\\
&+&\mu\sum_{l\sigma}C^{\dagger}_{l\sigma}C_{l\sigma}
+J\sum_{\langle ll'\rangle_{\rm NN}}{\bf S}_{l}\cdot {\bf S}_{l'},~~
\end{eqnarray}
where we consider only the nearest-neighbor (NN) and next NN hopping terms, the summations
$\langle ll'\rangle_{\rm NN}$ and $\langle ll'\rangle_{\rm NNN}$ denote that $l$ runs over all sites,
and for each $l$, over its NN sites and next NN sites, respectively, $C^{\dagger}_{l\sigma}$
($C_{l\sigma}$) is the creation (annihilation) operator for an electron of spin $\sigma$ on site $l$,
${\bf S}_{l}$ is a local spin operator, and $\mu$ is the chemical potential. This $t$-$J$ model
(\ref{tjham}) is subject to an important on-site local constraint to avoid the double electron
occupancy: $\sum_{\sigma}C^{\dagger}_{l\sigma}C_{l\sigma}\leq 1$. In order to satisfy this local
constraint, we employ the charge-spin separation fermion-spin formalism \cite{Feng15,Feng9404}, in
which the constrained electron operators $C_{l\uparrow}$ and $C_{l\downarrow}$ are replaced by,
\begin{eqnarray}\label{CSS}
C_{l\uparrow}=h^{\dagger}_{l\uparrow}S^{-}_{l}, ~~~~
C_{l\downarrow}=h^{\dagger}_{l\downarrow}S^{+}_{l},
\end{eqnarray}
respectively, where the spinful fermion operator $h_{l\sigma}=e^{-i\Phi_{l\sigma}}h_{l}$ keeps track
of the charge degree of freedom of the constrained electron together with some effects of spin
configuration rearrangements due to the presence of the doped hole itself (charge carrier), while
the spin operator $S_{l}$ represents the spin degree of freedom of the constrained electron, and
then the local constraint of no double occupancy is satisfied at each site in analytical calculations.
In this fermion-spin representation (\ref{CSS}), the original $t$-$J$ model (\ref{tjham}) can be
rewritten as,
\begin{eqnarray}\label{cssham}
H&=&t\sum_{\langle ll'\rangle_{\rm NN}}(h^{\dagger}_{l'\uparrow}h_{l\uparrow}S^{+}_{l}S^{-}_{l'} +h^{\dagger}_{l'\downarrow}h_{l\downarrow}S^{-}_{l}S^{+}_{l'}) \nonumber\\
&-&t'\sum_{\langle ll'\rangle_{\rm NNN}}(h^{\dagger}_{l'\uparrow}h_{l\uparrow}S^{+}_{l}S^{-}_{l'}
+h^{\dagger}_{l'\downarrow}h_{l\downarrow}S^{-}_{l}S^{+}_{l'}) \nonumber\\
&-&\mu\sum_{l\sigma}h^{\dagger}_{l\sigma}h_{l\sigma}+J_{{\rm eff}}\sum_{\langle ll'\rangle_{\rm NN}}
{\bf S}_{l}\cdot {\bf S}_{l'},
\end{eqnarray}
where $S^{-}_{l}=S^{\rm x}_{l}-iS^{\rm y}_{l}$ and $S^{+}_{l}=S^{\rm x}_{l}+iS^{\rm y}_{l}$ are the
spin-lowering and spin-raising operators for the spin $S=1/2$, respectively,
$J_{{\rm eff}}=(1-\delta)^{2}J$, and $\delta=\langle h^{\dagger}_{l\sigma}h_{l\sigma}\rangle$ is the
charge-carrier doping concentration. Based on the $t$-$J$ model in this fermion-spin representation
(\ref{cssham}), the kinetic-energy-driven SC mechanism has been established
\cite{Feng0306,Feng12,Feng15,Feng15a}, where the interaction between the charge carriers directly
from the kinetic energy of the $t$-$J$ model (\ref{cssham}) by the exchange of a strongly dispersive
{\it spin excitation} is responsible for the d-wave charge-carrier pairing in the particle-particle
channel, then the d-wave electron pairs originated from the d-wave charge-carrier pairing state are
due to the charge-spin recombination, and their condensation reveals the d-wave SC-state. The
characteristic features of the kinetic-energy-driven SC mechanism can be summarized as
\cite{Feng0306,Feng12,Feng15,Feng15a}: (i) the mechanism is purely electronic without phonons; (ii)
the mechanism indicates that the strong electron correlation favors superconductivity, since the main
ingredient is identified into an electron pairing mechanism not involving the phonon, the external
degree of freedom, but the internal spin degree of freedom of the constrained electron; (iii) the
electron pairing state is controlled by both the electron pair gap and single-particle coherence,
leading to that the maximal $T_{\rm c}$ occurs around the optimal doping, and then decreases in both
the underdoped and the overdoped regimes. Within the framework of this kinetic-energy-driven
superconductivity \cite{Feng0306,Feng12,Feng15,Feng15a}, the renormalization of the electrons in
cuprate superconductors has been investigated recently \cite{Gao18,Gao18a,Gao19}, and the obtained
main features of the single-particle excitation spectrum are well reproduced. The following analyses
of the normal and anomalous self-energies in cuprate superconductors build on this
kinetic-energy-driven SC mechanism \cite{Feng0306,Feng12,Feng15,Feng15a}. In these previous works
\cite{Feng15a}, the single-particle diagonal and off-diagonal propagators of the $t$-$J$ model in the
SC-state have been obtained in terms of the full charge-spin recombination, and can be expressed
explicitly as,
\begin{subequations}\label{EGF}
\begin{eqnarray}
G({\bf k},\omega)&=&{1\over\omega-\varepsilon_{\bf k}-\Sigma_{\rm tot}({\bf k},\omega)},\label{DEGF}\\
\Im^{\dagger}({\bf k},\omega)&=&{L_{\bf k}(\omega)\over \omega- \varepsilon_{\bf k}
-\Sigma_{\rm tot}({\bf k},\omega)},\label{ODEGF}
\end{eqnarray}
\end{subequations}
where the single-electron band energy $\varepsilon_{\bf k}=-4t\gamma_{\bf k}+4t'\gamma_{\bf k}'+\mu$,
with $\gamma_{\bf k}=({\rm cos}k_{x}+{\rm cos} k_{y})/2$, $\gamma_{\bf k}'={\rm cos}k_{x}{\rm cos}k_{y}$,
and $L_{\bf k}(\omega)=-\Sigma_{\rm pp}({\bf k},\omega)/[\omega+\varepsilon_{\bf k}
+\Sigma_{\rm ph}({\bf k},-\omega)]$, while the total self-energy $\Sigma_{\rm tot}({\bf k},\omega)$ is a
well-known combination of the normal self-energy $\Sigma_{\rm ph}({\bf k},\omega)$ and the anomalous
self-energy $\Sigma_{\rm pp} ({\bf k},\omega)$ as,
\begin{eqnarray}
\Sigma_{\rm tot}({\bf k},\omega)=\Sigma_{\rm ph}({\bf k},\omega)+W_{\bf k}(\omega), ~~~~~~\label{TOT-SE}
\end{eqnarray}
with the additional contribution $W_{\bf k}(\omega)$ below $T_{\rm c}$ due to the SC gap opening,
\begin{eqnarray}
W_{\bf k}(\omega)={|\Sigma_{\rm pp}({\bf k},\omega)|^{2}\over \omega+\varepsilon_{\bf k}
+\Sigma_{\rm ph}({\bf k}, -\omega)}.~~\label{TOT-MIX}
\end{eqnarray}
In the framework of the kinetic-energy-driven superconductivity, both the normal and anomalous
self-energies $\Sigma_{\rm ph}({\bf k},\omega)$ and $\Sigma_{\rm pp}({\bf k},\omega)$ arise from the
interaction between electrons mediated by a strongly dispersive spin excitation, and have been derived
explicitly in Ref. \onlinecite{Feng15a}, where all order parameters and chemical potential are
determined by the self-consistent calculation. In this sense, our calculation for both the normal and
anomalous self-energies is controllable without using adjustable parameters. In particular, the sharp
peak visible for temperature $T\rightarrow 0$ in the normal (anomalous) self-energy is actually a
$\delta$-functions, broadened by a small damping used in the numerical calculation at a finite lattice.
The calculation in this paper for the normal (anomalous) self-energy is performed numerically on a
$160\times 160$ lattice in momentum space, with the infinitesimal $i0_{+}\rightarrow i\Gamma$ replaced
by a small damping $\Gamma=0.1J$.

The single-particle spectral function $A({\bf k},\omega)$ measured by ARPES experiments is related
directly to the imaginary part of the single-particle diagonal propagator in Eq. (\ref{DEGF}) as
\cite{Damascelli03,Campuzano04,Fink07},
\begin{eqnarray}\label{ESF}
A({\bf k},\omega)={-2{\rm Im}\Sigma_{\rm tot}({\bf k},\omega)\over [\omega-\varepsilon_{\bf k}
-{\rm Re}\Sigma_{\rm tot}({\bf k},\omega)]^{2}+[{\rm Im} \Sigma_{\rm tot}({\bf k},\omega)]^{2}},~~~
\end{eqnarray}
where ${\rm Re}\Sigma_{\rm tot}({\bf k},\omega)$ and ${\rm Im}\Sigma_{\rm tot}({\bf k},\omega)$ are
the real and imaginary parts of the total self-energy $\Sigma_{\rm tot}({\bf k},\omega)$, respectively.
In ARPES experiments \cite{Damascelli03,Campuzano04,Fink07}, the energy renormalization of the
electrons in cuprate superconductors is directly determined by the real part of the total self-energy,
while the lifetime renormalization of the electrons is completely governed by the imaginary part of
the total self-energy. This is also why only the total self-energy can be extracted directly from
ARPES experiments. In the following discussions, the parameters in the $t$-$J$ model are chosen as
$t/J=3.5$ and $t'/t=0.4$. However, when necessary to compare with the experimental data, we take
$J=100$ meV, which is the typical value of cuprate superconductors
\cite{Damascelli03,Campuzano04,Fink07}.

\section{Quantitative characteristics}
\label{Quantitative-characteristics}

\begin{figure}[h!]
\centering
\includegraphics[scale=0.80]{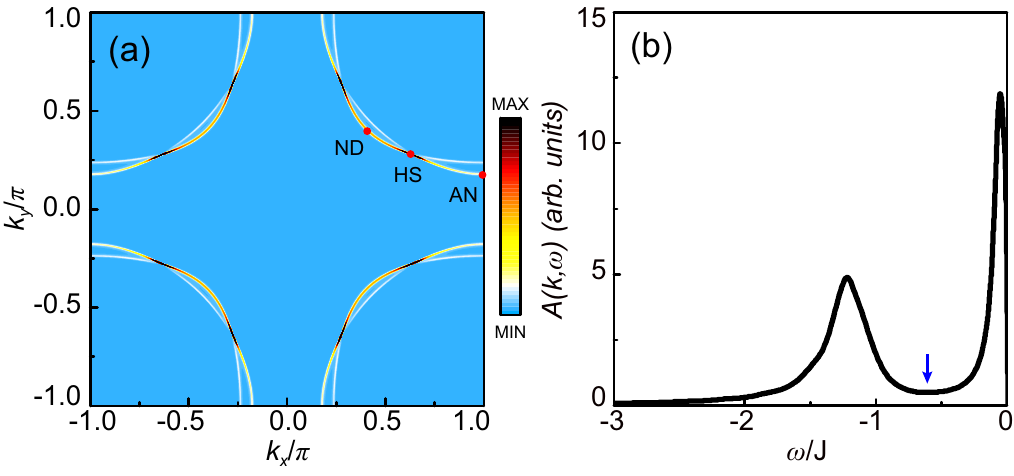}
\caption{(Color online) (a) The electron Fermi surface map and (b) the single-particle excitation
spectrum at the antinode as a function of energy in $\delta=0.15$ with $T=0.002J$ for $t/J=3.5$
and $t'/t=0.4$, where AN, HS, and ND in (a) denote the antinode, tip of the Fermi arc, and node,
respectively, while the blue arrow in (b) indicates the position of the dip. \label{EFS-PDH}}
\end{figure}

In an interacting electron system, the topology of the electron Fermi surface (EFS) plays a crucial
role in the understanding of the physical properties, since everything happens near EFS. In
particular, the strong coupling between the electrons and a strongly dispersive spin excitation in
cuprate superconductors leads to a strong redistribution of the spectral weights on EFS
\cite{Gao18,Gao18a,Gao19}, which cause the normal and anomalous self-energies to strongly vary with
the Fermi angle around EFS. For a convenience in the following discussions, (a) the underlying EFS
map \cite{Gao18} and (b) the single-particle excitation spectrum \cite{Gao18,Gao18a}
$A({\bf k}_{\rm AN},\omega)$ at the antinode as a function of energy for doping $\delta=0.15$ with
temperature $T=0.002J$ are {\it replotted} in Fig. \ref{EFS-PDH}. The result in Fig. \ref{EFS-PDH}a
therefore shows that EFS has been separated into three characteristic regions due to the strong
redistribution of the spectral weight: (a) the antinodal region, where the spectral weight is
suppressed, leading to that EFS around the antinodal region becomes unobservable in experiments
\cite{Norman98,Kanigel06,Yoshida06,Kondo13}; (b) the nodal region, where the spectral weight is
reduced modestly, leading to that EFS is truncated to form the disconnected Fermi arcs located around
the nodal region \cite{Norman98,Kanigel06,Yoshida06,Kondo13}; (c) the region at around the tips of
the Fermi arcs, where the renormalization from the quasiparticle scattering further reduces almost
all spectral weight on Fermi arcs to the tips of the Fermi arcs
\cite{Loret18,Chatterjee06,He14,Sassa11}. In this case, the spectral intensity exhibits a largest
value at around the tips of the Fermi arcs, where the characteristic feature is that both the real
and imaginary parts of the normal self-energy have the anomalously small values \cite{Gao18,Gao18a}.
In particular, the Fermi arcs collapse for the number of lattice sites $N\rightarrow\infty$ at
$T\rightarrow 0$, leading to form the Fermi-arc-tip liquid. Moreover, these tips of the Fermi arcs
connected by the scattering wave vectors ${\bf q}_{i}$ construct an {\it octet} scattering model,
which is a basic scattering model in the explanation of the Fourier transform scanning tunneling
spectroscopy (STS) experimental data, and also can give a consistent description of the regions of
the highest joint density of states detected from ARPES autocorrelation experiments
\cite{Gao19,Chatterjee06}. On the other hand, the result in Fig. \ref{EFS-PDH}b indicates that the
characteristic feature in the single-particle excitation spectrum is the dramatic change in the
spectral line-shape \cite{Dessau91,Norman97,Campuzano99,Wei08,DMou17}, where a quasiparticle peak
develops at the lowest energy, followed by a dip and a hump, giving rise to a striking PDH structure.
All these theoretical results are well consistent with the corresponding results observed from the
ARPES experiments.

\begin{figure}[h!]
\centering
\includegraphics[scale=0.85]{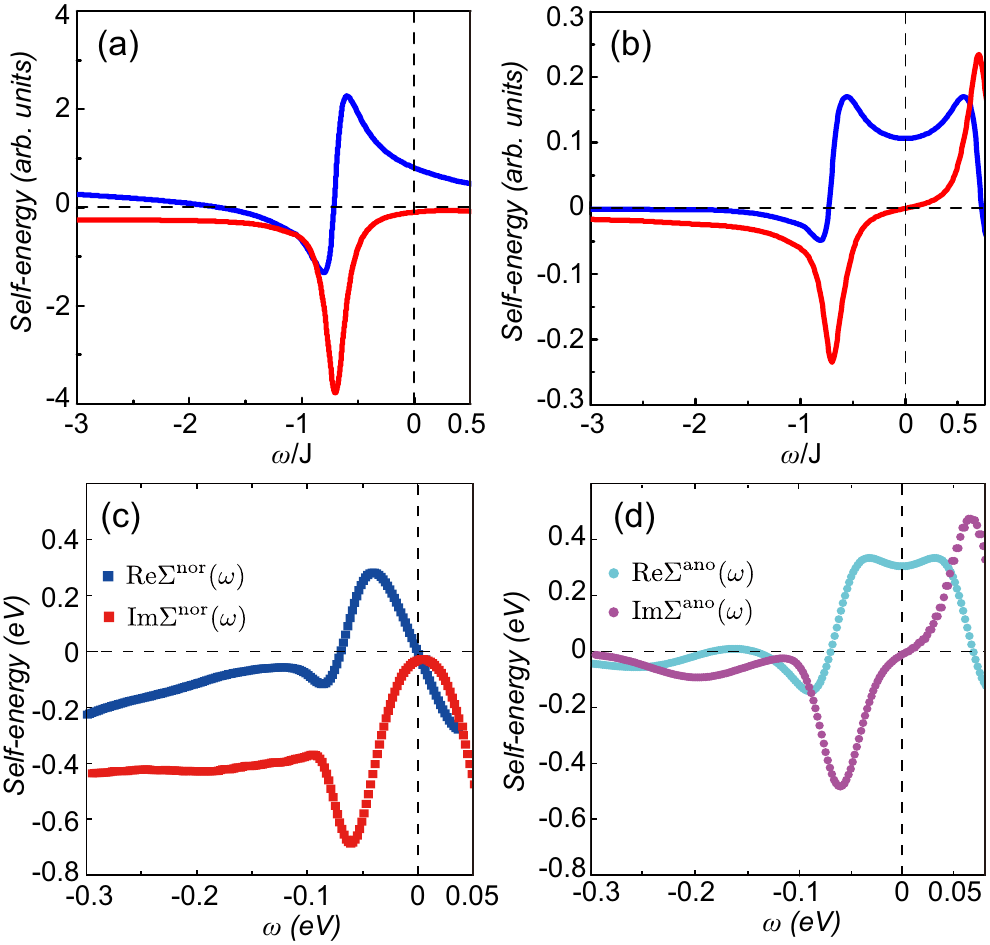}
\caption{(Color online) (a) The real (blue line) and imaginary (red line) parts of the normal
self-energy and (b) the real (blue line) and imaginary (red line) parts of the anomalous self-energy
at the antinode as a function of energy in $\delta=0.15$ with $T=0.002J$ for $t/J=3.5$ and $t'/t=0.4$.
The corresponding results of (c) the real and imaginary parts of the normal self-energy and (d) the
real and imaginary parts of the anomalous self-energy around the antinode as a function of energy
deduced from the ARPES spectra of the optimally doped Bi$_{2}$Sr$_{2}$CaCu$_{2}$O$_{8+\delta}$ via
the machine learning taken from Ref. \onlinecite{Yamaji19}. \label{self-energy-AN}}
\end{figure}

We are now ready to analyze the doping and momentum dependence of the normal and anomalous self-energies
in cuprate superconductors. In Fig. \ref{self-energy-AN}, we plot (a) the real (blue line) and imaginary
(red line) parts of the normal self-energy and (b) the real (blue line) and imaginary (red line) parts
of the anomalous self-energy at the antinode as a function of energy for $\delta=0.15$ with $T=0.002J$.
For a better comparison, the corresponding results \cite{Yamaji19} of (c) the real and imaginary parts
of the normal self-energy and (d) the real and imaginary parts of the anomalous self-energy around the
antinode as a function of energy deduced from the ARPES spectra of the optimally doped
Bi$_{2}$Sr$_{2}$CaCu$_{2}$O$_{8+\delta}$ via the machine learning technique are also shown in
Fig. \ref{self-energy-AN}. Apparently, the main low-energy features of both the normal and anomalous
self-energies deduced via the machine learning technique \cite{Yamaji19} are qualitatively reproduced,
where all the real and imaginary parts of the normal and anomalous self-energies exhibit the prominent
peak-structures in the low-energy region. Since the strength of the electron pair is directly determined
by the anomalous self-energy, these peaks in both ${\rm Re}\Sigma_{\rm pp}({\bf k}_{\rm AN},\omega)$ and
${\rm Im}\Sigma_{\rm pp}({\bf k}_{\rm AN},\omega)$ give rise to a crucial contribution to the SC gap,
and therefore are the true origin of the high $T_{\rm c}$ in cuprate superconductors \cite{Yamaji19}. On
the other hand, since the single-particle coherence is associated directly with the normal self-energy,
these peaks in ${\rm Re}\Sigma_{\rm ph}({\bf k}_{\rm AN}, \omega)$ and
${\rm Im}\Sigma_{\rm ph}({\bf k}_{\rm AN},\omega)$ dominate the energy and lifetime renormalization of
the electrons, respectively. In particular, the sharp change with energy in
${\rm Re}\Sigma_{\rm pp}({\bf k}_{\rm AN},\omega)$ and the large damping in
${\rm Im}\Sigma_{\rm pp}({\bf k}_{\rm AN},\omega)$ shown in Fig. \ref{self-energy-AN}b are also consistent
with these results obtained in the very early numerical solution of the Eliashberg equations based on the
spin-polaron $t$-$J$ model \cite{Plakida97}, where the d-wave SC-state is mediated by the exchange of spin
fluctuations. Moreover, it was shown \cite{Onufrieva09,Monthoux92,Onufrieva02,Kancharla08,Onufrieva12}
that the single-particle excitation spectra in the spin resonance mode mediated SC-state are close to
those observed experimentally in cuprate superconductors, where both the normal and anomalous
self-energies exhibit the sharp low-energy peak-structures \cite{Onufrieva09}.

However, there is a substantial difference between theory and machine learning in the high-energy region,
namely, the weaker features in the real and imaginary parts of the normal and anomalous self-energies
occur in the high-energy region, while the calculation anticipates a flat featureless with the values of
both the real and imaginary parts of the normal self-energy that approach zero. However, the actual range
of the low-energy peak structures of the normal and anomalous self-energies is very similar in theory and
machine learning. In particular, the sharp peak in ${\rm Im}\Sigma_{\rm pp}({\bf k}_{\rm AN},\omega)$
locates at the same energy $\omega_{\rm Im-Th}\sim -70$ meV as that in
${\rm Im}\Sigma_{\rm ph}({\bf k}_{\rm AN},\omega)$, which has been confirmed by the deduced result of the
normal (anomalous) self-energy based on the machine learning approach \cite{Yamaji19}. Moreover, this
anticipated peak energy $\omega_{\rm Im-Th}\sim -70$ meV in the optimal doping is also qualitatively
consistent with the corresponding result \cite{Yamaji19} of $\omega_{\rm Im-ML}\sim -65$ meV deduced in
the optimally doped Bi$_{2}$Sr$_{2}$CaCu$_{2}$O$_{8+\delta}$ via the machine learning technique.

\begin{figure}[h!]
\centering
\includegraphics[scale=1.10]{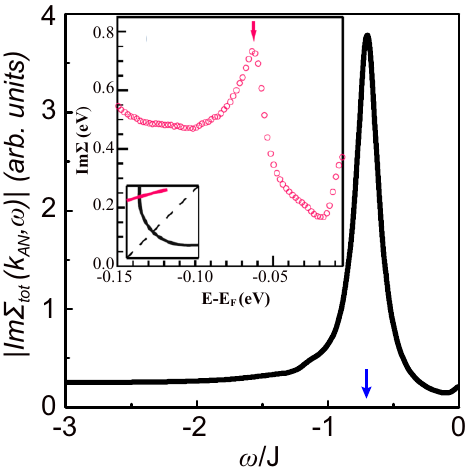}
\caption{(Color online) The imaginary part of the total self-energy at the antinode as a function of
energy in $\delta=0.15$ with $T=0.002J$ for $t/J=3.5$ and $t'/t=0.4$, where the blue arrow indicates
the position of the peak. Inset: the corresponding experimental result of the optimally doped
Bi$_{2}$Sr$_{2}$CaCu$_{2}$O$_{8+\delta}$ taken from Ref. \onlinecite{DMou17}. \label{scattering-rate}}
\end{figure}

In the SC-state, although the normal and anomalous self-energies describe theoretically different parts
of the interaction effects, all of them make the contributions to the dramatic change in the spectral
line-shape of the single-particle excitation spectrum. To see this point more clearly, we plot the
imaginary part of the total self-energy ${\rm Im}\Sigma_{\rm tot}({\bf k}_{\rm AN},\omega)$ [then the
quasiparticle scattering rate
$\Gamma({\bf k}_{\rm AN},\omega)=-{\rm Im}\Sigma_{\rm tot}({\bf k}_{\rm AN},\omega)$] at the antinode
as a function of energy for $\delta=0.15$ with $T=0.002J$ in Fig. \ref{scattering-rate} in comparison
with the corresponding ARPES experimental result \cite{DMou17} found in the optimally doped
Bi$_{2}$Sr$_{2}$CaCu$_{2}$O$_{8+\delta}$ around the antinode (inset). It should be noted that the
sharp low-energy peak-structure in ${\rm Im} \Sigma_{\rm tot}({\bf k}_{\rm AN},\omega)$ in
Fig. \ref{scattering-rate} is inconsistent with the corresponding result \cite{Yamaji19} deduced from
the optimally doped Bi$_{2}$Sr$_{2}$CaCu$_{2}$O$_{8+\delta}$ via the machine learning technique, where
although the sharp low-energy peaks appear in both the normal and anomalous self-energies, they cancel
in the imaginary part of the total self-energy to make the structure apparently invisible. However,
this sharp low-energy peak-structure in ${\rm Im}\Sigma_{\rm tot}({\bf k}_{\rm AN},\omega)$ in
Fig. \ref{scattering-rate} is very well consistent with the corresponding experimental result observed
\cite{DMou17} in the optimally doped Bi$_{2}$Sr$_{2}$CaCu$_{2}$O$_{8+\delta}$. At the antinode,
${\rm Im}\Sigma_{\rm tot}({\bf k}_{\rm AN},\omega)$ reaches a sharp peak at the energy of $-70$ meV,
and then the weight of the peak decreases rapidly in both the low-energy and high-energy ranges.
Concomitantly, this theoretical peak-energy of $-70$ meV in the optimum doping is also in qualitatively
agreement with the peak energy \cite{DMou17} of $-62$ meV detected in the optimally doped
Bi$_{2}$Sr$_{2}$CaCu$_{2}$O$_{8+\delta}$. More surprisedly, the position of this sharp low-energy
peak in ${\rm Im}\Sigma_{\rm tot}({\bf k}_{\rm AN},\omega)$ is just corresponding to the position of
the dip in the PDH structure in the single-particle excitation spectrum shown in Fig. \ref{EFS-PDH}b,
and therefore the sharp low-energy peak-structure in
${\rm Im}\Sigma_{\rm tot}({\bf k}_{\rm AN},\omega)$ induces an intensity depletion in the
single-particle excitation spectrum around the dip \cite{Gao18a}. The physical origin of this intensity
depletion around the dip in the PDH structure [then the sharp low-energy peak-structure in
${\rm Im}\Sigma_{\rm tot}({\bf k},\omega)$] can be attributed to the emergence of the momentum
dependence of the pseudogap. This follows a fact that the normal self-energy
$\Sigma_{\rm ph}({\bf k},\omega)$ in Eq. (\ref{EGF}) can be also rewritten as \cite{Feng15a},
\begin{eqnarray}\label{EPG}
\Sigma_{\rm ph}({\bf k},\omega)\approx {[\bar{\Delta}_{\rm PG}({\bf k})]^{2}\over\omega
+\varepsilon_{0{\bf k}}},
\end{eqnarray}
where the corresponding energy spectrum $\varepsilon_{0{\bf k}}$ and the pseudogap
$\bar{\Delta}_{\rm PG}({\bf k})$ are derived directly from the normal self-energy
$\Sigma_{\rm ph}({\bf k},\omega)$ and its antisymmetric part $\Sigma_{\rm pho} ({\bf k},\omega)$ as
$\varepsilon_{0{\bf k}}=-\Sigma_{\rm ph}({\bf k},0)/\Sigma_{\rm pho}({\bf k},0)$ and
$\bar{\Delta}_{\rm PG}({\bf k})=\Sigma_{\rm ph}({\bf k},0)/\sqrt{-\Sigma_{\rm pho}({\bf k},0)}$,
respectively, and have been given explicitly in Ref. \onlinecite{Feng15a}. This pseudogap
$\bar{\Delta}_{\rm PG}({\bf k})$ is therefore identified as being a region of the electron self-energy
effect by which it means the spectral intensity is suppressed. In particular, the result in
Eq. (\ref{EPG}) also indicates that the imaginary part of the total self-energy
${\rm Im}\Sigma_{\rm tot}({\bf k},\omega)\propto {\rm Im}\Sigma_{\rm ph}({\bf k},\omega)\approx
2\pi[\bar{\Delta}_{\rm PG}({\bf k})]^{2}\delta(\omega+\varepsilon_{0{\bf k}})$, and then the
pseudogap $\bar{\Delta}_{\rm PG}({\bf k})$ plays the same role in the inducement of an intensity
depletion in the single-particle excitation spectrum around the dip as that of
${\rm Im}\Sigma_{\rm tot}({\bf k},\omega)$. In other words, the pseudogap-induced low-energy
peak-structure in ${\rm Im}\Sigma_{\rm tot}({\bf k},\omega)$ [then $\bar{\Delta}_{\rm PG}({\bf k})$]
in Fig. \ref{scattering-rate} is directly responsible for the famous PDH structure in the
single-particle excitation spectrum shown in Fig. \ref{EFS-PDH}b. Moreover, since
${\rm Im}\Sigma_{\rm tot}({\bf k},\omega)\propto {\rm Im}\Sigma_{\rm ph}({\bf k},\omega)
\sim [\bar{\Delta}_{\rm PG}({\bf k})]^{2}$, the appearance of the sharp low-energy peak-structure in
${\rm Im}\Sigma_{\rm tot}({\bf k},\omega)$ is determined mainly by
${\rm Im}\Sigma_{\rm ph}({\bf k},\omega)$ [then $\bar{\Delta}_{\rm PG}({\bf k})$], reflecting a basic
fact that the main feature of the pseudogap-induced low-energy peak-structure in
${\rm Im}\Sigma_{\rm tot}({\bf k}_{\rm AN},\omega)$ can persist into the normal-state \cite{Gao18a},
leading to that the PDH structure is totally unrelated to superconductivity.

\begin{figure}[h!]
\centering
\includegraphics[scale=0.83]{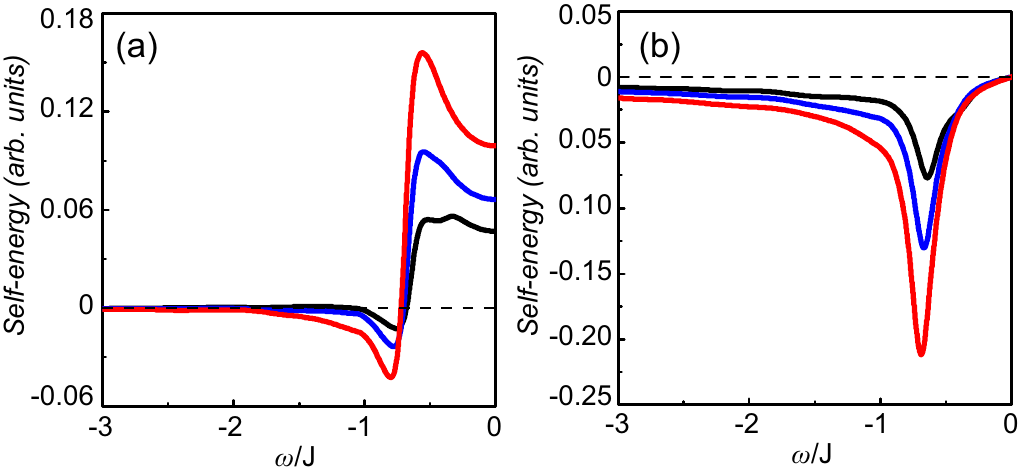}
\caption{(Color online) The (a) real and (b) imaginary parts of the anomalous self-energy at the
antinode as a function of energy in $\delta=0.06$ (black line), $\delta=0.09$ (blue line), and
$\delta=0.12$ (red line) with $T=0.002J$ for $t/J=3.5$ and $t'/t=0.4$.  \label{self-energy-AN-doping}}
\end{figure}

As a natural consequence of the doped Mott insulators, the normal and anomalous self energies in
cuprate superconductors evolve with doping. In Fig. \ref{self-energy-AN-doping}, we plot the results
of the (a) real and (b) imaginary parts of the anomalous self-energy at the antinode as a function of
energy for $\delta=0.06$ (black line), $\delta=0.09$ (blue line), and $\delta=0.12$ (red line) with
$T=0.002J$. It is thus shown clearly that in the underdoped regime, when the doping concentration is
increased, (i) the low-energy peaks in both ${\rm Re}\Sigma_{\rm pp}({\bf k}_{\rm AN},\omega)$ and
${\rm Im}\Sigma_{\rm pp}({\bf k}_{\rm AN},\omega)$ move further away from the Fermi energy, and (ii)
the weights of these low-energy peaks are increased, which are nothing, but the SC gap that increases
in magnitude with doping in the underdoped regime. Moreover, the evolution of the imaginary part of
the normal self-energy with doping at around the $[\pi,0]$ point of BZ has been also investigated
\cite{Gao18a}, and results show that in the underdoped regime, the low-energy peak in
${\rm Im}\Sigma_{\rm ph} ({\bf k},\omega)|_{{\bf k}=[\pi,0]}$ shifts further towards to the Fermi
energy with the increase of doping, which leads to that both the hump and lowest-energy peak in the PDH
structure of the single-particle excitation spectrum at around the $[\pi,0]$ point move further towards
to the Fermi energy with the increase of doping, also in qualitative agreement with the corresponding
ARPES experimental results \cite{Campuzano99}.

\begin{figure}[h!]
\centering
\includegraphics[scale=1.10]{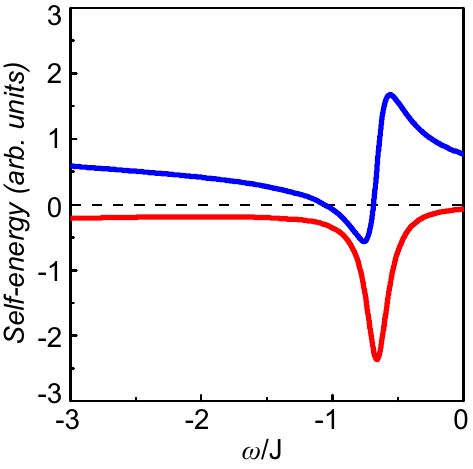}
\caption{(Color online) The real (blue line) and imaginary (red line) parts of the normal self-energy
at the node as a function of energy in $\delta=0.15$ with $T=0.002J$ for $t/J=3.5$ and $t'/t=0.4$.
\label{self-energy-ND}}
\end{figure}

Now we turn to discuss the evolution of the normal and anomalous self-energies with momentum. In
Fig. \ref{self-energy-ND}, we plot the real (blue line) and imaginary (red line) parts of the normal
self-energy at the node as a function of energy for $\delta=0.15$ with $T=0.002J$. In comparison with
the corresponding result in Fig. \ref{self-energy-AN}a for the same set of parameters except for at
the node, it is shown clearly when one moves the momentum ${\bf k}_{\rm F}$ from the antinode to the
node, the weights of the low-energy peaks in both ${\rm Re}\Sigma_{\rm ph}({\bf k}_{\rm ND},\omega)$
and ${\rm Im}\Sigma_{\rm ph}({\bf k}_{\rm ND},\omega)$ are reduced, while the low-energy peaks move
further towards to the Fermi energy. On the other hand, in the kinetic-energy-driven SC mechanism
\cite{Feng0306,Feng12,Feng15,Feng15a}, the characteristic feature of the d-wave SC-state is the
existence of four nodes on EFS, where the SC gap vanishes $\Sigma_{\rm pp}({\bf k_{\rm ND}},\omega)=0$.
Moreover, it has been found in the previous studies that the low-energy peak structures in both the
normal and anomalous self-energies disappear at around the tips of the Fermi arcs
\cite{Gao18,Gao18a,Gao19}, where $\Sigma_{\rm ph}({\bf k_{\rm HS}},\omega)$ and
$\Sigma_{\rm pp}({\bf k_{\rm HS}},\omega)$ have the anomalously small values, reflecting a fact that
the coupling strength of the electrons to a strongly dispersive spin excitation is quite weak
\cite{Mou19}. Concomitantly, the imaginary part of the total self-energy
${\rm Im}\Sigma_{\rm tot}({\bf k}_{\rm HS},\omega)$ has an anomalously small value at around the tips
of the Fermi arcs \cite{Gao18,Gao18a,Gao19}, which has been confirmed by the experiments
\cite{Loret18,Chatterjee06,He14,Sassa11}, where the weakest quasiparticle scattering that occurs at
around the tips of the Fermi arcs has been observed. This is also why the spectral intensity exhibits
a largest value at around the tips of the Fermi arcs shown in Fig. \ref{EFS-PDH}.

\begin{figure}[h!]
\centering
\includegraphics[scale=0.87]{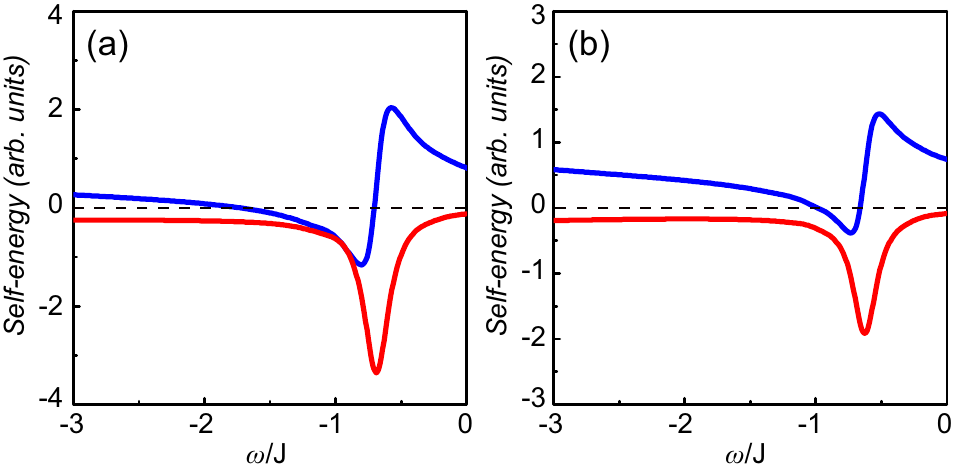}
\caption{(Color online) The real (blue line) and imaginary (red line) parts of the normal self-energy
at (a) the antinode and (b) the node as a function of energy in $\delta=0.15$ with $T=0.15J$ for
$t/J=3.5$ and $t'/t=0.4$. \label{self-energy-normal}}
\end{figure}

At the temperature above $T_{\rm c}$, the electrons are in a normal-state, where the SC gap
$\Sigma_{\rm pp}({\bf k},\omega)=0$. However, the low-energy peak-structure in the normal self-energy
can persist into the normal-state. To see this point more clearly, we plot the results of the real
(blue line) and imaginary (red line) parts of the normal self-energy at (a) the antinode and (b) the
node as a function of energy for $\delta=0.15$ with $T=0.15J$ in Fig. \ref{self-energy-normal}. In
comparison with the corresponding results in Fig. \ref{self-energy-AN}a and Fig. \ref{self-energy-ND}
for the same set of parameters except for in the normal-state ($T=0.15J$), one can find that although
the weights of the low-energy peaks are suppressed with the increase of temperatures, the positions of
these low-energy peaks in the normal-state do not change much from the corresponding case in the
SC-state. This is why some characteristic features in the single-particle excitation spectrum of cuprate
superconductors arising from the renormalization of the electrons can be detected from experiments in
both the SC-state and normal-state.

As we have mentioned above in Sec. \ref{Formalism}, one of the characteristic features in the
kinetic-energy-driven SC mechanism is that the SC-state is controlled by both the electron pair gap and
single-particle coherence. In this case, to examine the microscopic theory of the kinetic-energy-driven
superconductivity, it need to compare the obtained normal and anomalous self-energies with the
corresponding results deduced via the machine learning method, and the obtained total self-energy with
the corresponding experimental data. Our present study shows that in the kinetic-energy-driven
superconductivity, the exchanged spin excitations give rise to the low-energy peak-structures in both
the normal self-energy $\Sigma_{\rm ph}({\bf k},\omega)$ and anomalous self-energy
$\Sigma_{\rm pp}({\bf k},\omega)$, which are in qualitative agreement with the corresponding low-energy
peak-structures in both the normal and anomalous self-energies deduced from the ARPES spectra of
cuprate superconductors via the machine learning technique \cite{Yamaji19}. Concomitantly, these
low-energy peak-structures in both the normal and anomalous self-energies do not cancel in the total
self-energy, and then the well-pronounced low-energy peak-structure also appears in the total
self-energy $\Sigma_{\rm tot}({\bf k},\omega)$, which is well consistent with the corresponding
low-energy peak-structure in the total self-energy observed experimentally on cuprate superconductors
\cite{DMou17}. The qualitative agreement between the low-energy peak-structures in both the normal and
anomalous self-energies obtained based on the kinetic-energy-driven superconductivity and those
deduced from the ARPES spectra of cuprate superconductors via the machine learning technique
\cite{Yamaji19} together with the good agreement between theory and experiment \cite{DMou17} for the
low-energy peak-structure in the total self-energy therefore shows why the theory of the
kinetic-energy-driven superconductivity can give a consistent description of the renormalization of
the electrons in cuprate superconductors \cite{Gao18,Gao18a,Gao19}.

In the machine learning analysis in Ref. \onlinecite{Yamaji19}, the low-energy peak-structures in both
the hidden normal and anomalous self-energies of cuprate superconductors are deduced solely from the
complicated ARPES line-shape, which therefore provide a fingerprint of the SC mechanism. However,
within the present machine learning method \cite{Yamaji19}, the deduced cancellation effect of these
low-energy peaks in the total self-energy making the structure apparently invisible is inconsistent with
the corresponding experimental observations \cite{DMou17}, where the notable low-energy peak-structure
in the total self-energy has been observed experimentally. On the other hand, the strong coupling of the
electrons with a strongly dispersive spin excitation in cuprate superconductors induces a strong EFS
reconstruction, which complicate the low-energy electronic state properties. Our present study therefore
call for a systematic analysis with the improvements in the machine learning method to obtain the more
accurate results of the hidden quantities, including the normal and anomalous self-energies, at all
around EFS from the experimental data observed from the ARPES measurements with the improved resolution,
together with other powerful measurement techniques, such as STS, Raman scattering, and infrared
measurements of the reflectance. These more accurate results of the hidden quantities would be crucial
to the understanding of the essential physics of cuprate superconductors and the related
kinetic-energy-driven SC mechanism.

\section{Conclusions}\label{Conclusion}

In summary, we have compared the results of the normal and anomalous self-energies deduced from the ARPES
spectra of cuprate superconductors via the machine learning technique with these obtained based on the
kinetic-energy-driven superconductivity, and the obtained results show that both the normal and anomalous
self-energies due to the interaction between electrons mediated by a strongly dispersive spin excitation
exhibit the notable low-energy peak-structures at all around EFS except for at the tips of the Fermi arcs,
where the low-energy peak-structures are predicted to be absent \cite{Gao18,Gao18a,Gao19}. However, these
prominent low-energy peak-structures in both the normal and anomalous self-energies do not cancel in the
total self-energy, leading to the appearance of the low-energy peak-structure in the total self-energy.
In particular, the low-energy peak-structure in the normal self-energy is mainly responsible for the PDH
structure in the single-particle excitation spectrum, and can persist into the normal-state, while the
low-energy peak in the anomalous self-energy gives rise to a crucial contribution to the SC gap, and
vanishes in the normal-state. Furthermore, these low-energy peak-structures in both the normal and
anomalous self-energies evolve with doping, where the low-energy peak in the anomalous self-energy at
around the antinode region moves further away from the Fermi energy as the doping concentration is
increased in the underdoped regime. More specifically, these low-energy peaks have a special momentum
dependence, where although the weight of the low-energy peak in the normal self-energy is gradually reduced
when one moves the momentum from antinode to node, the position of the low-energy peak moves towards to the
Fermi energy.

%~~\\
\section*{Acknowledgements}

This work was supported by the National Key Research and Development Program of China under
Grant No. 2016YFA0300304, and the National Natural Science Foundation of China under Grant Nos. 11974051
and 11734002.

\end{document}